\documentclass[a4paper,11pt]{article}
\usepackage{jinstxxx}
\usepackage[utf8]{inputenc} 
\usepackage[T1]{fontenc}    
\usepackage{hyperref}       
\usepackage{url}            
\usepackage{booktabs}       
\usepackage{amsfonts}       
\usepackage{nicefrac}       
\usepackage{microtype}      
\usepackage{amsmath}
\usepackage{siunitx}
\usepackage{gensymb}
\usepackage{multirow}
\usepackage{graphicx}
\usepackage{textcomp} 
\usepackage{soul,color}          
\usepackage{chngcntr} 
\usepackage{lineno}

\usepackage{siunitx}
\usepackage{xspace}

\newcommand{\eg}{{\it e.g.}}

\newcommand{\micro}{\ensuremath{\mu}}

\newcommand{\bbonu}{\ensuremath{\beta\beta0\nu}}

\newcommand{\Tonu}{\ensuremath{T_{1/2}^{0\nu}}}

\newcommand{\SEVEN}{\ensuremath{\textbf{7}}}
\newcommand{\FPEAK}{\ensuremath{f_\lambda}}
\newcommand{\SEVENBA}{\textbf{7}$\cdot Ba^{2+}$}

\newcommand{\XE}{\ensuremath{{}^{136}\rm Xe}}
\newcommand{\GE}{\ensuremath{{}^{76}\rm Ge}}

\newcommand{\TE}{\ensuremath{{}^{128}\rm Te}}

\newcommand{\Bapp}{\ensuremath{{\rm Ba^{2+}}}}

\DeclareSIUnit\c{\mbox{$c$}}
\DeclareSIUnit\magn{\mbox{$\times$}}
\DeclareSIUnit\min{min}
\DeclareSIUnit\week{week}
\DeclareSIUnit\year{yr}
\DeclareSIUnit\years{years}
\DeclareSIUnit\yr{yr}
\DeclareSIUnit\standard{std}
\DeclareSIUnit\str{sr}
\DeclareSIUnit\ppm{ppm}
\DeclareSIUnit\ppb{ppb}
\DeclareSIUnit\ppt{ppt}
\DeclareSIUnit\pe{PE}
\DeclareSIUnit\spe{SPE}
\DeclareSIUnit\ev{events}
\DeclareSIUnit\ct{counts}
\DeclareSIUnit\neutron{\mbox{$n$}}
\DeclareSIUnit\smp{samples}
\DeclareSIUnit\Sample{S}
\DeclareSIUnit\ch{ch}
\DeclareSIUnit\hit{hit}
\DeclareSIUnit\hits{hits}
\DeclareSIUnit\bin{(\mbox{5-PE}~bin)}
\DeclareSIUnit\sgm{\mbox{$\sigma$}}
\DeclareSIUnit\rms{RMS}
\DeclareSIUnit\keVr{\mbox{keV$_{\rm nr}$}}
\DeclareSIUnit\keVee{\mbox{keV$_{e{\rm e}}$}}
\DeclareSIUnit\ph{photon}
\DeclareSIUnit\pes{pes}
\DeclareSIUnit\el{electrons}
\DeclareSIUnit\pm{PMT}
\DeclareSIUnit\inch{"}
\DeclareSIUnit\bit{bit}
\DeclareSIUnit\sample{samples}
\DeclareSIUnit\barn{barn}
\DeclareSIUnit\bara{bar}
\DeclareSIUnit\barg{barg}
\DeclareSIUnit\mlardepth{\mbox(meter~of~\LAr~depth)}
\DeclareSIUnit\Curie{Ci}
\DeclareSIUnit\psi{psi}
\DeclareSIUnit\parsec{pc}
\DeclareSIUnit\liveday{\mbox{live-days}}
\DeclareSIUnit\days{\mbox{days}}
\DeclareSIUnit\day{\mbox{day}}
\DeclareSIUnit\miles{\mbox{miles}}
\DeclareSIUnit\degreeC{\mbox{$^{\circ}$C}}
\DeclareSIUnit\electron{\mbox{$e^-$}}
\DeclareSIUnit\Euro{\mbox{\euro}}
\DeclareSIUnit\cph{cph}
\DeclareSIUnit\neq{neq}
\DeclareSIUnit\unit{unit}
\DeclareSIUnit\byte{Byte}
\DeclareSIUnit\Bq{\becquerel}

\begin{document}
\title{Towards a background-free neutrinoless double beta decay experiment based on a fluorescent bicolor sensor}

\author[a]{\small Iv\'an Rivilla,}

\author[b]{\small Borja Aparicio,}

\author[c]{\small Juan M.~ Bueno,}

\author[a, d]{\small David Casanova,}

\author[a]{\small Claire Tonnel\'e,}

\author[d, e]{\small Zoraida Freixa,}

\author[a]{\small Pablo Herrero,}

\author[f]{\small Jos\'e I.~ Miranda,}

\author[c]{\small Rosa M.~ Mart\'inez-Ojeda,}

\author[a, d]{\small Francesc Monrabal,}

\author[g]{\small Be\~nat  Olave,}

\author[g, d]{\small Thomas Sch\"{a}fer,}

\author[c]{\small Pablo Artal,}

\author[h]{\small David Nygren,}

\author[a, b, 1]{\small Fernando P.~ Coss\'io,}

\author[a, d, 1]{\small Juan J.~ G\'omez-Cadenas}

\note{corresponding author}

\emailAdd{jjgomezcadenas@dipc.org}

\emailAdd{fp.cossio@ehu.es}

\affiliation[a]{\footnotesize Donostia International Physics Center (DIPC), Manuel Lardizabal Ibilbidea 4, 20018 San Sebasti\'an / Donostia, Spain}
\affiliation[b]{\footnotesize Department of Organic Chemistry I, University of the Basque Country (UPV/EHU), Centro de Innovaci\'on en Qu\'imica Avanzada (ORFEO-CINQA), Manuel Lardizabal Ibilbidea 3, 20018 San Sebasti\'an / Donostia, Spain}
\affiliation[c]{\footnotesize Laboratorio de \'Optica (LOUM) \&
Centro de Investigaci\'on en \'Optica y Nanof\'isica (CiOyN), University of Murcia, Espinardo Campus, 
30100 Murcia, Spain}
\affiliation[d]{\footnotesize Ikerbasque, Basque Foundation for Science, Mar\'ia D\'iaz de Haro 3, 6, 48013 Bilbao, Spain}
\affiliation[e]{\footnotesize Department of Applied Chemistry, Faculty of Chemistry, University of the Basque Country (UPV-EHU), San Sebasti\'an / Donostia, Spain}
\affiliation[f]{\footnotesize SGIker NMR Facility, University of the Basque Country (UPV/EHU), Avda. Tolosa 72, E-20018 San Sebasti\'an / Donostia, Spain}
\affiliation[g]{\footnotesize NanoBioSeparations Group, POLYMAT, University of the Basque Country (UPV/EHU), Avda. Tolosa 72, E-20018 Donostia/San Sebasti\'an, Spain}
\affiliation[h]{\footnotesize Department of Physics, University of Texas at Arlington, Arlington, TX 76019, USA}

\abstract{
Searching for neutrinoless double beta decays (\bbonu) is the only practical way to establish if the neutrinos are their own antiparticles. Due to the smallness of neutrino masses, the lifetime of \bbonu\ is expected to be at least ten orders of magnitude smaller than the noise associated with the natural radioactive chains. A positive identification of \bbonu\ decays requires, ultimately, finding a signal that cannot be mimicked by radioactive backgrounds. This signal could be the observation of the daughter atom in the decay, since no known background processes induce a Z+2 transformation. In particular,  the \bbonu\ decay of \XE\ could be established by detecting the doubly ionised daughter atom, \Bapp. Such a detection could be achieved via a sensor made of a monolayer of molecular indicators. The \Bapp\ would be captured by one of the molecules in the sensor, and the presence of the single chelated indicator would be subsequently revealed by a strong fluorescent response from repeated interrogation with a laser system. 
Here we describe a fluorescent bicolor indicator that binds strongly to \Bapp\ 
and shines very brightly, shifting its emission colour from green to blue when chelated in dry medium, thus allowing the unambiguous identification of single barium atoms in the sensor, and permitting a positive identification of the \bbonu\ decay of \XE\ in a gas chamber, that could led to a background-free experiment.}

\notoctrue
\maketitle
\flushbottom
\notoc


\section{Introduction}

Neutrinoless double beta decay (\bbonu) is a hypothetical, very slow radioactive process in which two neutrons 
undergo $\beta$-decay simultaneously and without the emission of neutrinos, 
$(Z,A) \rightarrow (Z+2,A) + 2\ e^{-}$.  An unambiguous observation of this process would establish that neutrinos are Majorana particles \cite{Majorana:1937}, identical to their antiparticles. A discovery would have deep implications in particle physics and cosmology \cite{Fukugita:1986hr}.

Double beta decay (DBD) experiments have been searching \bbonu\ decay in several isotopes for more than half a century without finding clear evidence of a signal. The current best lower limit on the lifetime (\Tonu) of \bbonu\ processes has been obtained for the isotope \XE, for which
$\Tonu > 10^{26}$ \si{\year} \cite{Gando:2016ji}. Two other isotopes, \GE\ and \TE, have also been studied with similar results \cite{Agostini:2018tnm, Alduino:2017ehq}. A new generation of \bbonu\ experiments will aim to improve the sensitivity to \Tonu\ by several orders of magnitude, in order to maximise the chances of
a discovery \cite{Gomez-Cadenas:2019sfa}. This will require very large exposures, measured in ton-years, but even more importantly, 
a greatly enhanced capability to suppress the background associated with 
natural radioactivity and  cosmogenic activation. 

The most powerful discriminant against backgrounds would be detection of the daughter atom in a double beta decay, since no known background processes induce a Z+2 transformation.  The possibility of barium tagging in a xenon TPC was proposed in 1991 by Moe \cite{Moe:1991ik}, and has been extensively investigated for the last two decades \cite{Danilov:2000pp, Sinclair:2011zz,  Mong:2014iya}. Recently
the nEXO collaboration has demonstrated the imaging and counting of individual barium atoms in solid xenon by scanning a focused laser across a solid xenon matrix deposited on a sapphire window \cite{Chambers:2018srx}. This is a promising step for barium tagging in liquid xenon. The technique originally proposed by Moe and being pursued by nEXO relies in Ba$^+$ fluorescence imaging using two atomic excitation levels in very low density gas. In liquid xenon, recombination is frequent and the barium daughters are distributed across charge states from 0 to 2+ \cite{PhysRevC.92.045504}, with sizeable populations of neutral Ba and  Ba$^+$.  In the high pressure gas phase, however, recombination is minimal \cite{1997NIMPA.396..360B}, and \Bapp\ dominates. 

In 2015, Nygren proposed a \Bapp\ sensor based on a monolayer of fluorescent molecular indicators that could be incorporated into high-pressure gas xenon  (HPXe) time projection chambers (TPCs) \cite{Nygren_2015}, such as those being developed by the 
NEXT Collaboration \cite{Nygren:2009zz, Alvarez:2012haa, Martin-Albo:2015rhw, Gomez-Cadenas:2019sfa}. The concept that was further developed in \cite{Jones:2016qiq}. The NEXT detectors already provide two strong handles for background suppression, namely, excellent energy resolution \cite{Renner:2019pfe}, 
and the capability to reconstruct the combined trajectory of the two electrons emitted in the 
decay \cite{Ferrario:2019kwg}. In Nygren's proposal, the observation of the \Bapp\ ion produced in the \bbonu\ decay could be associated in time and space with the appearance of the double electron signature. 

The technique envisioned the transport {\it in situ} of the doubly charged ion to the TPC cathode.  Theoretical calculations indicate that \Bapp\ will drift in various solvation states to the cathode \cite{Bainglass:2018odn}. Once near the cathode, the ion must be focused, for example using RF \cite{ARAI201456}, into a small region where the sensor will be located. Alternatively, the sensor can be installed in a moving system that can intercept the ion when it reaches the cathode plane. This notion relies on the fact that the reconstruction of the electrons in the chamber allows the prediction of the arrival position at the cathode of the \Bapp\ within a radius of a few \si{\cm} and within a time window of less than
\SI{1}{\milli\second}. 

A molecule whose response to optical stimulation changes when it forms a supramolecular complex with a specific ion is a fluorescent indicator, and ions thus bound to molecules are generally referred to as being chelated. In the case of a HPXe TPC for a \bbonu\ search, the indicator must be designed to chelate \Bapp\ ions with high efficiency. To detect the chelated molecule, the sensor is scanned with a suitable laser system and the chelated molecule identified by its characteristic response. The molecule must not only capture the ion but must then fluoresce at a solid-gas interface in a dry medium. In contrast, most common indicators used in biochemistry rely in chelating groups such as polyamines and polycarboxylates, which undergo reorganisation of the chelating group-metal ion complex in solution. In dry media \cite{Jeong:2012, Carter:2014}, the necessary reorganisation is frustrated, making these unsuitable for chemical interactions in ultra-dry solid-gas interfaces.  Recently, indicators able to fluoresce in dry media have been developed in the context of  R\&D performed by the NEXT collaboration to develop a more suitable \Bapp\ sensor \cite{Thapa:2019zjk}. 

Nygren's proposal was followed by an initial proof of concept which resolved individual 
\Bapp\ ions on a scanning surface \cite{McDonald:2017izm}. The sensor was a thin quartz plate with surface-bound fluorescent indicators continuously illuminated with excitation light and monitored by an EM-CCD camera to record the emission light. The  indicator 
was Fluo-3 
suspended in polyvinyl alcohol (PVA) to immobilise the molecular complex and facilitate optical imaging. In order to excite only a thin layer above the quartz-PVA interface, total internal reflection (TIRF) illumination was used. The results of \cite{McDonald:2017izm}  demonstrated single-ion sensitivity through detection exceeding 12 standard deviations, confirmed by single-step photobleaching, and constituted en essential first step toward barium tagging in a HPXe.

And yet, the sensor of \cite{McDonald:2017izm}, could not operate in a HPXe. The gas in a xenon TPC must be free of impurities such as water and alcohols at a level that would rule out a PVA-based sensor as well as Fluo-3, which is heavily quenched in dry media. In addition, TIRF was achieved by coupling the microscope's objective to the sensor via standard optical oil, but the use of such oil in the HPXe TPC \bbonu\ detector would be problematic. Furthermore, the sensor in an experiment implementing \Bapp\ sensitivity will need to be densely populated by indicators to ensure maximum ion-capture efficiency, while the target of \cite{McDonald:2017izm} was sparse in order to facilitate the observation of individual chelated molecules. In a densely packed monolayer, a single chelated molecule may be surrounded by as many as $10^6$ unchelated molecules per squared micron and thus a discrimination factor as large as possible is desirable to achieve the large signal-to-noise-ratio needed for an unambiguous identification of the \Bapp\ in a reasonable scanning time.  


In this paper we report the development of a fluorescent bicolor indicator (FBI) synthesised to bind strongly to \Bapp\ 
and to shine very brightly when complexed with \Bapp\ in dry medium, so that chelated molecules emit $\sim$300 times more light than their unchelated siblings. Furthermore, the emission spectrum of the chelated indicators is significantly blue-shifted  with respect to the unchelated species, allowing a robust separation of both spectra that provides and additional discrimination factor of 40. As a consequence, the discrimination factor between the chelated and unchelated species in dry medium is found to be $1.2 \times 10^4$, more than three orders of magnitude larger than that found for common indicators like Fluo-3. 
 We then use two photon absorption (TPA) microscopy to provide a proof of concept of a technique leading to unambiguous identification of \Bapp\ ions in a sensor based on FBI indicators.


\section{Design, synthesis and characterisation of FBI compounds}


Our criteria to design FBI are summarised in Figure 1. The indicator is an  ensemble of 
four components (Figure  1a). A metal binding group, a fluorophore, a photophysically inert spacer and, finally, a linker to a substrate. Figure  1b illustrates the desired behaviour of the indicator upon chelation with \Bapp\ ions. We require that: {\bf a}, the chelating group binds the cation with a high binding constant; {\bf b}, the indicator response in dry medium is preserved and preferably enhanced w.r.t. the response in solution, and {\bf c}, the fluorophore exhibits distinct response in the visible region for the chelated and unchelated states (thus the term bicolor indicator). To that end, the synthesis of FBI compounds incorporates a custom-designed fluorophore, based on nitrogen-containing aromatic polyheterocycles \cite{Aginagalde:2010, Zhang:2018, Ko:2018, Maitra:2017}. The shift in response upon coordination provides a strong signature of a chelated indicator, which exhibits a blue shift, over a background of unchelated species. Furthermore we require that the indicator response does not form supra-molecular complexes with the light elements in the barium column (e.g, beryllium, calcium and magnesium). This eliminates a possible source of background, since those light elements are abundant in the environment, although notably, likely not in their 2$^+$ charge states in xenon gas.


\begin{table}[bth!]
\begin{center}
  \begin{tabular}{|c|c|c|c|c|c|c|c|}
    \hline
    \multirow{1}{*}{Cmpd.} &
      \multicolumn{2}{|c|}{a} & 
      \multicolumn{1}{|c|}{b} &
      \multicolumn{2}{|c|}{c} &
       \multicolumn{2}{|c|}{d} \\ \hline
         &   \multicolumn{2}{|c|}{$\lambda_{em} (nm)$} & 
      \multicolumn{1}{|c|}{$f_\lambda$} &
      \multicolumn{2}{|c|}{$\Phi_\lambda$} &
       \multicolumn{2}{|c|}{$B_\lambda(M^{-1} cm^{-1})$} \\ \hline
    & \SEVEN & \SEVENBA & \SEVENBA & \SEVEN &  \SEVENBA & \SEVEN & \SEVENBA \\
    \hline
    \textbf{7aa} &	485 &	485 &	0.07	& 0.42 &	0.41	& 8.42 &	8.45 \\
    \hline
\textbf{7ba}	&482	 &428 &	6.02	& 0.34 &	0.32 &	7.65 &	8.13\\
\hline
\textbf{7ca	} &489 &428& 	179.74 &	0.67	&0.45 &	11.26&	8.06 \\
\hline
\textbf{7da}	&491	&491	& n. d. &	0.06	&0.06 &	0.53	& 0.51 \\
\hline
\textbf{7ec	} &511	&430	 &22.64 &	0.29	& 0.25 &	3.65	&3.05 \\
\hline
\textbf{7cb} 	&503	 &456 & 4.86 &	0.22	& 0.04	&4.84 &	1.21\\
\hline
  \end{tabular}
  \end{center}
\caption{Characterisation of FBI compounds \SEVEN\ and \SEVENBA. {\bf a}, Emission wavelengths at an excitation wavelength of \SI{250}{nm}. {\bf b}, Peak discrimination factors, \FPEAK, with respect to unbound fluorophores \textbf{7} at $\lambda_{em}$. {\bf c}, Quantum yields,  $\Phi_\lambda$, at $\lambda_{em}$.  {\bf d}, Molecular brightnesses of the fluorescent emissions, $B_\lambda$, at $\lambda_{em}$; n. d.: not determined.}
\end{table}

The chemical synthesis of our sensors is shown in Figure 1c. The process starts with the double 
addition-elimination reaction between 2-aminopyridines (X=CH) \textbf{1a-c} or 2-aminopyrimidine \textbf{1b} (X=N) and 2,4-dibromoacetophenone \textbf{2}. Bicyclic heterocycles \textbf{3a-c} reacted with aza-crown ethers \textbf{4a-c} in the presence of a Pd(0)/DavePhos catalytic system to generate intermediates \textbf{5a-e} in moderate (30\%) to very good (95\%) yields. Finally, these latter adducts were coupled with aromatic 1,2-dibromides \textbf{6a,b} by means of a catalytic system formed by a Pd(II) salt and XPhos to yield the desired FBI compounds \textbf{7aa-cb}.
In this latter step, the formal (8+2) reaction was carried out in the presence of cesium carbonate as a weak base when R=ethoxycarbonyl (compound \textbf{7ec}) in order to prevent decarboxylation of the ester group. 


Our experiments to determine the photo-physical properties of compounds \textbf{7} started by recording their respective emission spectra in acetonitrile solutions. Although all compounds were fluorescent with large intensities in the minimum energy transitions, the critical criterion to select the most suitable candidate was the ability of a given compound to exhibit different lowest emission wavelengths in their unbound and barium-coordinated forms. We defined the peak discrimination factor \FPEAK\ at a given wavelength $\lambda$ as:

\begin{equation}
f_\lambda = \frac{I_\lambda (7 \cdot Ba^{2+}) - I_\lambda (7)}{I_\lambda (7)}
\label{eq:discFactor}             
\end{equation}
where $I_\lambda (7 \cdot Ba^{2+})$ and $I_\lambda (7)$ are the intensities of the emission signals at wavelength $\lambda$ of the corresponding bound (\SEVENBA) and free (\SEVEN) fluorophore, respectively. In addition, we measured the molecular brightness \cite{Carter:2014} $B_\lambda$ of each transition according to the following expression:

\begin{equation}
B_\lambda = \epsilon_\lambda \phi_\lambda    
\label{eq:bright}       
\end{equation}
where $\epsilon_\lambda$ is the molar extinction coefficient and $\phi_\lambda$ is the emission quantum yield. 

The data associated with the photophysics of compounds \textbf{7} are collected in Table 1. According to our results, compound \textbf{7aa}, possessing the 1,4,7-trioxa-10-azacyclododecane moiety (\textbf{4a}, n=1) does not show any significant difference between the free and barium-bound states, thus indicating that this four-heteroatom aza-crown ether is too small to accommodate the $Ba^{2+}$ cation. Compound \textbf{7ba}, with a 1,4,7,10-tetraoxa-13-azacyclopentadecane unit (\textbf{4b}, n=2) showed a noticeable blue shift upon coordination with $Ba^{2+}$, ($\Delta\lambda=-54$ nm). However, the low value of $f_\lambda$ makes this size of the chelating group non optimal for further developments. In the case of FBI molecule \textbf{7ca}, which incorporates the six-heteroatom-containing aza-crown ether unit 1,4,7,10,13-pentaoxa-16-azacyclooctadecane (\textbf{4c}, n=3), a larger blue shift associated with $Ba^{2+}$ coordination  ($\Delta\lambda=-61$ nm) was observed. Most importantly, the $f_\lambda$ discrimination factor was found to be of ca. 180, which shows a significant separation between the unbound \textbf{7ca} and $Ba^{2+}$-coordinated \textbf{7ca}-\Bapp\ species. Both emission spectra are gathered in Figure \ref{fig:emission}. 
In addition, both unbound and cationic species showed acceptable quantum yields and molecular brightness values. 

Having  selected compound \textbf{7ca} as the best FBI candidate, we conducted studies 
to assess its binding ability, which must be high (in dry medium) for our sensor. To that end, we measured first its cation association constant $K_a$ with barium perchlorate in acetonitrile at 298 K by means of the Benesi-Hildebrand method \cite{Benesi:1949} and the corresponding fluorescence spectra, according to the following formula \cite{ZhangDuan:2018}:

\begin{equation}
\frac{1}{F - F_{min}} = \frac{1}{F_{max} - F_{min}} \left(1+ \frac{1}{K_a [Ba^{2+}]} \right)
\label{eq:ka}
\end{equation}

In this expression, $F$ is the measured emission of compound \textbf{7ca} at the excitation wavelength $\lambda_{exc} = 250$ nm in the presence of a given [$Ba^{2+}$] concentration, whereas $F_{min}$ and $F_{max}$ stand for the corresponding intensities of free aza-crown ether \textbf{7ca} and host-guest complex \textbf{7ca$\cdot Ba^{2+}$}, respectively. Under these conditions and on the basis of the data shown in Figure \ref{fig:emission}d, we measured a binding constant of 
${K_a}$=5.26 \,10${^4}$ M$^{-1}$ (r$^{2}$=0.909). This indicates a very efficient ability of compound \textbf{7ca} for $Ba^{2+}$ capture and formation of the  (\textbf{7ca}$\cdot Ba^{2+}$)(ClO$_4{^-})_2$ salt in solution, whose favourable photophysical parameters are gathered in Table 1. In addition, the Job's plot showed a maximum for $n = m = 1$, thus indicating that \textbf{7ca} captures only one $Ba^{2+}$ cation per molecule, as it is shown in Figure  \ref{fig:emission}e.

As far as the chemical structure of the tetracyclic fluorophore is concerned, our results indicate that introducing an additional nitrogen heteroatom in the 2,2a$^{1}$-diazacyclopenta[jk]fluorene  to form the corresponding 2,2a$^{1}$,3-triazacyclopenta[jk]fluorene analogue is detrimental in terms of quantum yield and molecular brightness, as it can be appreciated from the photophysical properties of compound \textbf{7da} shown in Table 1. Moreover, the presence of an additional fused phenyl group in the fluorophore results in the formation of imidazo[5,1,2-cd]naphtho[2,3-a]indolizine derivative \textbf{7cb}, whose $f_\lambda$ factor was significantly lower than that measured for \textbf{7ca}. Therefore, the presence of additional fused aromatic or hetheroaromatic rings to the basic benzo[a]imidazo[5,1,2-cd]indolizine scaffold does not improve the photophysical properties of the resulting cycloadduct. Finally, the presence of an electron-withdrawing group in compound \textbf{7ec} results in a quenching of quantum yield of the fluorophore as well as a lowering of the discrimination factor. Consequently, we determine that further chemical elaboration of the fluorophore skeleton in order to synthesise the spacer and linker groups shown in Figure 1a, must not involve carboxy derivatives like esters or amides, but ${\pi}$-decoupled moieties such as alkoxy groups. Therefore we conclude that \textbf{7ca} is the optimal combination of structural and electronic features to fulfil our previously defined design criteria.

\section{Electronic structure calculations and NMR experiments}

Electronic structure calculations at the Density Functional Theory (DFT)  level both in the gas phase and in solution confirm the strong binding affinity of \textbf{7ca} to coordinate \Bapp. The \textbf{7ca}$\cdot$\Bapp\ optimised structure exhibits a large molecular torsion of the binding group with respect to the free \textbf{7ca} molecule (see the $\omega$ dihedral angle in 
Figure \ref{fig:nmr}b, calculation done at $\omega$B97X-D/6-311++G(p,d)\&Lanl2DZ level of theory), so that a molecular cavity appears, with the metal cation forming a $\pi$-complex between the \Bapp\ metallic centre and the phenyl group. The oxygen atoms of the aza-crown ether occupy five coordination positions with O$\cdots$Ba contacts within the range of the sum of the ionic radii (2.8-3.0 \AA) \cite{Batsanov:2001}. Interestingly, the phenyl ring attached to the crown ether is oriented towards the centre of the cavity coordinating \Bapp\ through the $\pi$-electrons. Frontier molecular orbitals (MO) of \textbf{7ca} are delocalised over the entire fluorophore moiety, with virtually no participation of the binding group electrons (Figure \ref{fig:nmr}c, computed at the $\omega$B97X-D/6-311G(d,p)/LANDL2DZ level). The lowest bright state of the unbound FBI molecule can be mainly characterised as the electronic transition between highest occupied MO (HOMO) and the lowest unoccupied MO (LUMO). Molecular distortion upon metal coordination in \textbf{7ca}$\cdot$\Bapp\ has an important impact on the electronic structure. In particular, the torsion of the phenyl group allowing $\pi$-coordination breaks the planarity with the rest of the fluorophore, modifying HOMO and LUMO energy levels. The decrease of the effective conjugation with respect to \textbf{7ca} increases the symmetry allowed ${\pi} \rightarrow {\pi}^*$ gap, thus resulting in the blue shift of the fluorescent emission (Figure \ref{fig:nmr}c). Therefore, these results support the viability of \textbf{7ca} as an efficient \Bapp\ indicator in both wet and dry conditions (see Supporting Information).

NMR Experiments on the complexation reaction between FIB molecule \textbf{7ca} and barium perchlorate are compatible with the geometries obtained by the DFT calculations. Progressive addition of the salt promoted a deshielding to lower field of the \textit{b} protons, which are in \textit{ortho} disposition with respect to the aza-crown ether. The \textit{meta} protons marked as \textit{c} in Figure \ref{fig:nmr}d showed a similar, but lower in magnitude, deshielding effect. The remaining protons of the benzo[a]imidazo[5,1,2-cd]indolizine fluorophore showed a very light deshielding effect, but remained essentially unchanged. Instead, the 1,4,7,10,13-pentaoxa-16-azacyclooctadecaane moiety of \textbf{7ca} showed different deshielding effects upon coordination with \Bapp\, with the only exception of the \textit{N}-methylene protons denoted as \textit{a} in Figure \ref{fig:nmr}e, which were shifted to higher field, thus demonstrating that the nitrogen atom of the aza-crown ether is not participating in the coordination with the dication.

\section{Characterisation of FBI in dry media}


To measure the response of FBI in dry media, we first studied the behaviour of this molecule embedded in different polymers. 
Specifically,  three polymers were studied as a support matrix: polyvinyl alcohol (PVA), polytmethyl metacrylate (PMMA) and
poly ether blockamide (PEBAX\textregistered\ 2533). 

Our results are illustrated in Figure \ref{fig:poly}a for PMMA. Under excitation light of 
\SI{350}{nm}, the spectra of both chelated and unchelated molecules are rather similar and cannot be effectively separated. 
Similar results are obtained for other excitation wavelengths. All the other polymers exhibit a similar behaviour. We attribute the lack of separation between the spectra of chelated and unchelated indicators to the restriction of the conformational freedom imposed by the polymer's rigid 
environment. 

A better alternative is provided by silica gel as a solid phase support. Adsorption of the molecule on the silica surface  permits the exposition of at least one side of its crown ether moiety to the interaction with \Bapp\ cations. In addition, this solid-gas interface topology preserves the conformational freedom required to reach the coordination pattern observed in our calculations. Therefore, the emission spectrum recorded on silica for the coordinated indicator keeps the essential features observed in solution, particularly the blue shift discussed above (Figure  \ref{fig:poly}c).

Two samples were manufactured. Sample SF was prepared depositing FBI ({\bf 7ca}) on silica gel in powdery form (\SI{104}{\milli\gram} of silica).  Sample SBF was formed depositing FBI on silica gel saturated with barium (SBF).
The concentration of sample SF was 
\SI{2.3E-5}{\milli\mol} of indicator per mg of silica, while the concentration of sample SBF was 
\SI{7.4E-8}{\milli\mol}  of indicator per mg of silica. The ratio $C_r$, between the concentrations of SF 
and SBF was $C_r = 307$. 

Figure \ref{fig:poly}b shows the spectra of both samples, upon irradiation at \SI{250}{nm}  after subtracting the fluorescence emitted by the silica. The spectra keep the same essential features (in particular the strong colour shift between the chelated and unchelated species) observed in solution (Figure \ref{fig:emission}a). On the other hand, the SBF samples have a concentration $\sim 300$ times smaller than that of the SF samples, implying, since the total area of both samples is similar, a brightness $\sim 300$ times larger. 

In a real experiment, we want to separate the blue-shifted light emitted by the chelated indicator from the green light emitted by the surrounding unchelated molecule. The simplest way to achieve this is to pass the light emitted by the sample through a low-pass filter, with a cutoff frequency, $\lambda_f$, chosen to maximise the signal to noise ratio between both species. Figure \ref{fig:poly}b shows the large separation achieved between the SF and SBF spectra when a cutoff at
\SI{450}{nm} (dashed black line) is set. 

Consider a sample containing 
a concentration of chelated and unchelated molecules given respectively by $C_{sbf}$ and $C_{sf}$, with ratio
$C_r = C_{sbf}/C_{sf}$. Call
 the fraction of the SBF spectrum below the cutoff, 
$f_{sbf} = SBF_{\lambda < \lambda_f} /SBF$, and analogously, define $f_{sf} = SF_{\lambda < \lambda_f} /SF$. We take as optimal value of the cutoff the wavelength that 
maximises the ratio $D_r = f_{sbf} / f_{sf}$ which has a broad maximum in the range \SIrange{440}{450}{nm}. Setting the cutoff at \SI{450}{nm}, we obtain $D_r = 40 \pm 4$. The light accepted by the filter for the chelated and unchelated species is, respectively 
$I_{sbf} = f_{sbf} \cdot C_{sbf}$ and $I_{sf} = f_{sf} \cdot C_{sf}$.  We define the global discrimination factor, $F$, between the chelated and unchelated spectra as  
\begin{equation}
F = I_{sbf}/I_{sf} = D_r \cdot C_r
\label{eq:f}
\end{equation}

For the SBF and SF samples we find
$F = (1.2 \pm 0.2)  \times 10^4$, where the 20\% error is obtained propagating the relative errors in the measurements of the spectra and the concentrations of the samples.

\section{Characterisation of FBI using two photon absorption microscopy}

Two-photon absorption (TPA) is the physical process by which  two photons are absorbed simultaneously by a molecule. The energy difference between the involved lower and upper states of the molecule equals  the sum of the energies of the two photons absorbed. Because TPA is a second-order process, the number of photons absorbed per molecule per unit time is proportional to the square of the incident intensity, which in turn is proportional to the beam power $P$. The number of absorbed photons, $n_a$, per fluorophere and per pulse, when the beam is focused in a diffraction-limited spot is \cite{Denk990}:

\begin{equation}
n_a = \frac{P^2 \delta}{\tau f} (\frac{A^2}{2 \hbar c \lambda})^2
\label{eq:na}
\end{equation}
where $P$ is the laser power, $A$ is the numerical aperture, $\delta$ is the fluorophore brightness ($\sigma \cdot \phi_\lambda$) of the fluorophore, $\tau$ the width of the
laser pulse and $f$ its repetition rate. 

Equation \ref{eq:na} highlights the basic requirements for TPA microscopy. Since $\delta$ is small (\eg\ 
$\delta = 36 \pm 9.7 $ GM for fluorescein, where 
1 GM$= 10^{-50}$ cm$^4 \cdot$second$/$ (photon$\cdot$molecule)), the laser setup must provide high power
(to exploit the non linear-term $P^2$), short-pulses and high repetition rate, so that the product
$\tau \cdot f$ is large. In addition, the numerical aperture must be as large as possible. Notice, however, that in dry medium, $A$ is limited, in practice, to a maximum value of 1, since no optical oil can be used to couple the microscope objective to the sample.
Moreover, notice that the maximum number of absorbed photons per pulse in TPA is two. Consequently, if the microscopy setup is able to compensate the low cross-section so that $n_a = 2$, the number of emitted fluorescent photons equals the laser rate. 

Compared with single-photon absorption, TPA has the advantage of providing self-focusing of the laser in a spot whose size is given by the diffraction limit. Given the impossibility to use optical oil in gas detectors, TIRF becomes difficult to achieve in a HPXe, and TPA microscopy appears as an excellent alternative for the laser scanning of the sensor. 

We have used a TPA microscopy setup \cite{Avila2019}  (see Methods for a detailed description) to study chelated and unchelated FBI indicators in dry medium. To that end, FBI ({\bf 7ca}) was deposited in silica gel in powdery form, and then compressed to form silica pellets suitable for laser scan. Sample SFp was formed depositing FBI in a pellet of \SI{35}{\milli\gram} of silica, while
sample SBFp was formed depositing FBI in a pellet of the same mass saturated with barium salt. The concentration of FBI in pellet SFp was identical to that prepared for FBI in powdery form (sample SF), and equal to \SI{2.3E-5}{\milli\mol} of indicator per mg of silica. Analogously, the concentration of SBFp was the same than that of SBF, and equal to 
\SI{7.4E-8}{\milli\mol}  of indicator per mg of silica. 
To  emulate the conditions of a real experiment, we have performed a laser scan using the SFp and SBFp pellets as models of dry sensors. Low- and high-pass filters were used to set the cutoff wavelength at
\SI{450}{nm}. 
A blank silica pellet was measured to determined the response of the substrate without the indicator, which was only appreciable at a very high power (\SI{240}{\milli\watt}). The SFp and SBFp pellets were very bright at much lower power (\SI{40}{\milli\watt}). While we subtract the signal of the blank from the response of the SFp and SBFp, its contribution is very small, of the order of 0.5\%.  

Figure  \ref{fig:laser} shows four representative XZ tomographic images or profiles. In all the cases, the laser moves in steps of \SI{2}{\micro\meter}, scanning the sample across the transverse coordinate $Z$ (\eg\, across the thickness of the pellet), and one longitudinal coordinate ($X$). Each point represented in the profiles corresponds to a volume of roughly 1 $\mu m^3$. Notice that the scans reveal the details of the deposited samples.

Panels {\bf a} and {\bf b} in Figure  \ref{fig:laser}, correspond to scans on the SBFp pellet. A high-pass \SI{450}{nm} filter was applied in panel {\bf a} (green panel) and a low-pass \SI{450}{nm} filter was applied in panel {\bf b} (blue panel). Panels {\bf c} and {\bf d} correspond to scans on the SFp pellet applying the same filters.  The fraction of light below \SI{450}{nm} is significant for the SBFp pellet  (23.2\% of the total) and very small for the SFp pellet (0.55\% of the total). We find that $D_r = 42$, in good agreement with the value found in powder, thus confirming our previous result and demonstrating the capability of TPA laser microscopy to separate efficiently chelated and unchelated FBI indicators.

\section{Discussion: towards a sensor for \Bapp\ tagging}


Here we collect our previous results to show that the intense brightness and large separation factor of FBI permits a robust observation of single chelated molecules even for densely packed sensors.

As an example consider a TPA microscopy system similar to the one used in this work, but with optimised parameters, e.g, $f = 100$ MHz, $\tau = 100$ fs and $A = 0.95$. Using a state-of-the art CCD camera with quantum efficiency of 65\%, the overall light collection efficiency of the system will be $\epsilon = 20$\%.  Focusing the laser in \SI{1}{\square\micro\meter} results in a photon density of \SI{1.9e+31}{photons \per pulse \per cm^2 \per \watt}. 

Assume now that a given FBI molecule has captured a \Bapp\ ion, and that the laser is focused in a \SI{1}{\square\micro\meter} spot containing this complexed indicator. We can compute the number of  photons that the chelated indicator absorbs as a function of the laser power, using eq. \ref{eq:na}. Given the relatively large TPA cross section of FBI (see Methods),  $n_a =2$ for a modest power of \SI{20}{mW}. Setting the laser power at this value, the emission rate of the chelated molecule will equal the laser repetition rate, \SI{1E+8}{photons\per\second} (see Figure \ref{fig:na}).   

The light emitted by the complexed FBI molecule will be blue-shifted. Assume that a low pass filter with a cutoff at 
$\lambda < 450$ nm is placed in front of the CCD. Call the fluorescence emitted in a given time interval by the chelated indicator $n_f$. The fraction of the fluorescence emitted by the unchelated molecules in the spot ($m -1 \sim m$) that will pass the filter is $n_b = \frac{n_f \,m}{F}$, where $F$ is the discrimination factor between chelated and unchelated species. Call $N_f = \epsilon \cdot n_f $ and $N_b = \epsilon \cdot n_b $ to the {\em detected} photons due to the chelated indicator and to the
unchelated molecules. The total signal recorded in the CCD will be $N_t = N_f + N_b$. The estimator of the signal observed in the spot will be $N_t - N_b$, where $N_b$ can be computed with great precision taking the average of a large number of spots containing only unchelated molecules. The  (SNR) of the subtraction is:

\begin{equation}
SNR = \frac{N_f}{\sqrt{N_b}} = \sqrt{\epsilon \frac{n_f\,F}{m}} = 
\sqrt{2 \times 10^{-1} \frac{10^8 \times 1.2 \times 10^4}{10^6}} = 490 \sqrt{\si{\second}}
\label{eq.snr}
\end{equation}

Thus for \SI{10}{ms}, $SNR \sim 50$. Therefore, a chelated indicator would produce an unmistakable signal above the background of unchelated molecules in that spot. This demonstrates that fast and unambiguous identification of 
\Bapp\ ions in the sensor can be attained using a dense monolayer and without resorting to single-molecule photobleaching. 

%

\clearpage
\section*{Figures}

\begin{figure}[htb!]
\begin{center}
\includegraphics[width=0.99\textwidth]{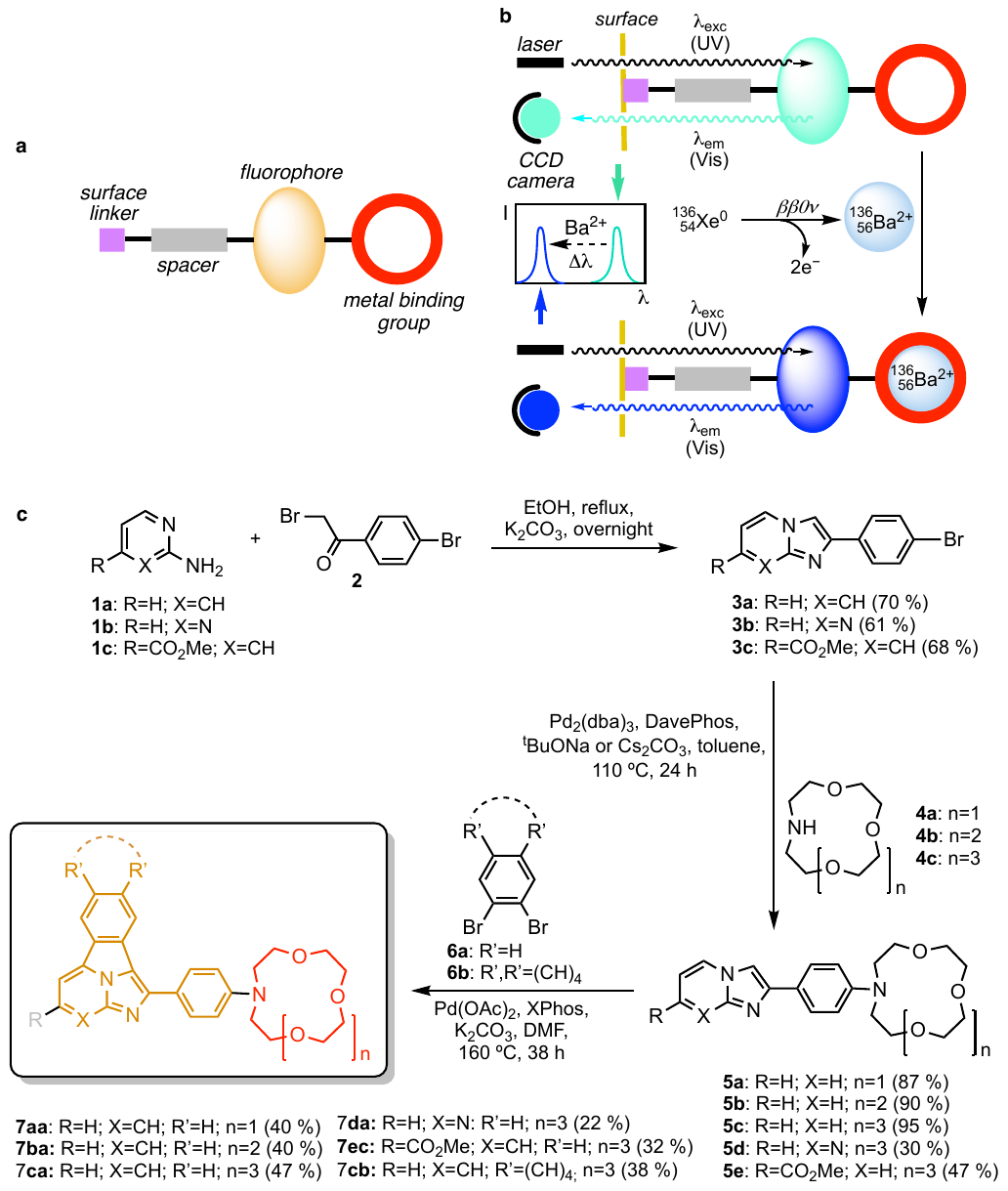}
\caption{{\it Design and synthesis of FBI}. The top panel shows the design criteria. {\bf a}, The components of the indicator and {\bf b}, the photophysical requirements. The bottom panel, {\bf c}, shows the chemical synthesis from  pyridines (or pyrimidines), bromoacetophenones, 1,2-dibromoarenes and aza-crown ethers.}
\end{center}
\label{fibDesign}
\end{figure}

\begin{figure}[htb!]
\centering
\includegraphics[width=0.60\textwidth]{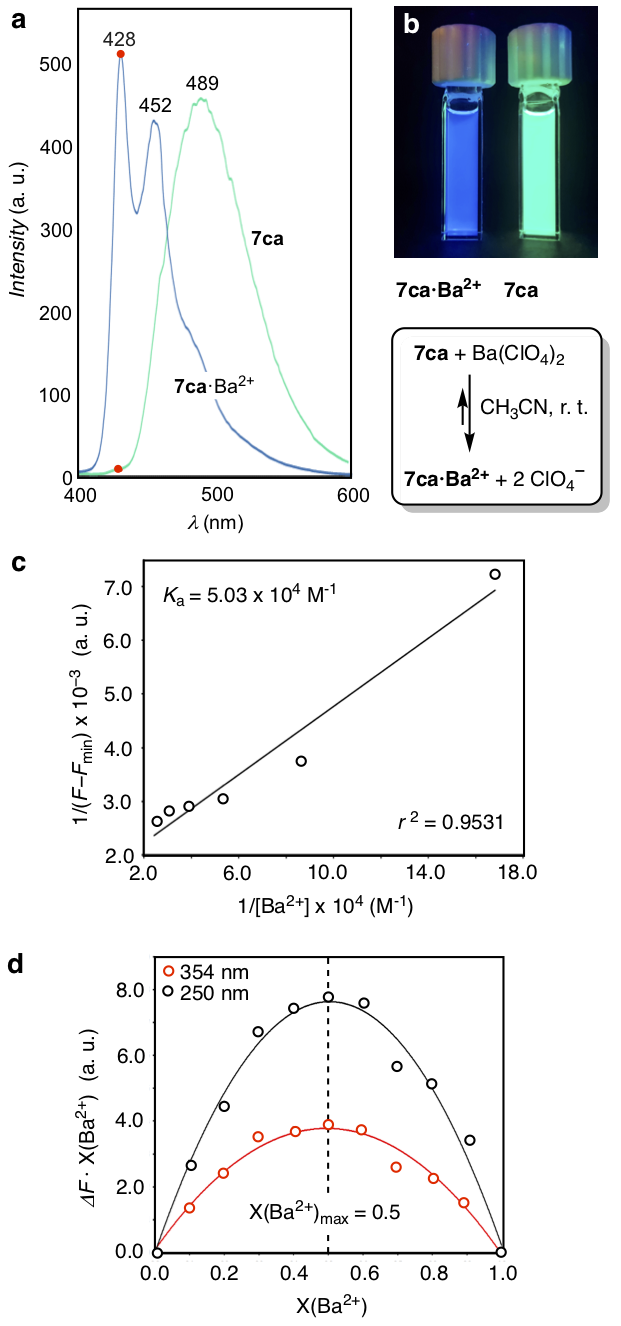} 
\caption{{\it Characterisation of FBI}. {\bf a}, Emission spectra of unchelated (\textbf{7ca}) and chelated (\textbf{7ca}$\cdot$\Bapp) indicators upon excitation at \SI{250}{nm}. Red dots indicate the wavelengths used to determine the peak  discrimination factor $f_\lambda$. {\bf b}, Photographs of both species in acetonitrile showing the bicolor emission, upon irradiation at \SI{365}{nm}. {\bf c}, Benesi-Hildebrand plot  of the fluorescence emission spectra of FBI in acetonitrile solution at room temperature in the presence of different concentrations of barium perchlorate. {\bf e}, Job's plot of the  \textbf{7ca}-Ba(ClO${_4}$)${_2}$ interaction showing a 1:1 stoichiometry between \textbf{7ca} and \Bapp\, thus forming complex \textbf{7ca}$\cdot$\Bapp.}
\label{fig:emission}
\end{figure}

\begin{figure}[htb!]
\centering
\includegraphics[width=0.8\textwidth]{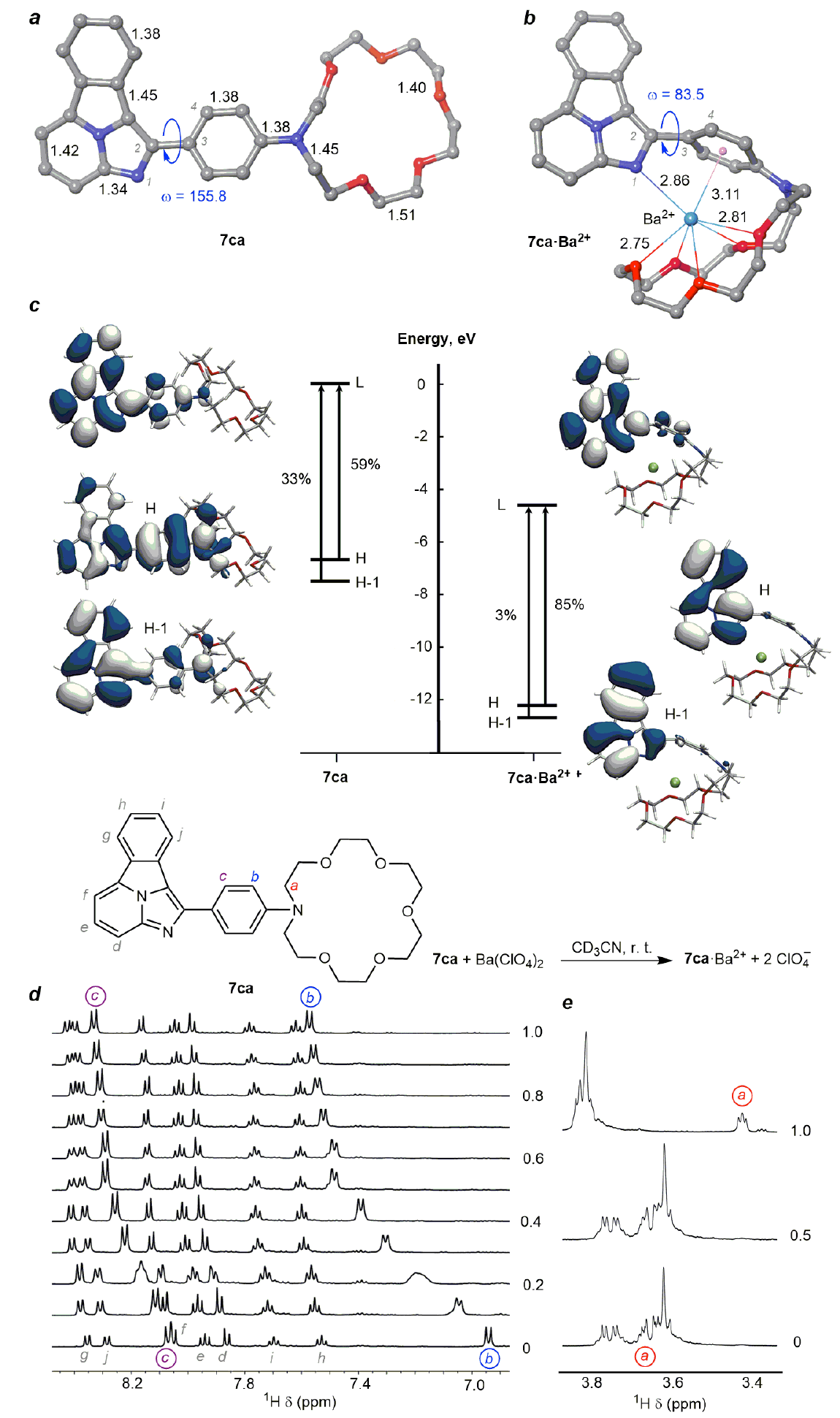} 
\caption{{\it Theoretical predictions and NMR experiments}. Top panel: Density functional theory gas phase structures of  {\bf a}, \textbf{7ca} and {\bf b}, \textbf{7ca}$\cdot$ \Bapp. Bond distances are given in \AA. Dihedral angles $\omega$ formed by covalently bonded atoms 1-4 are given in deg and in absolute value. {\bf c}, Frontier molecular orbital energy diagram of \textbf{7ca} (left) and \textbf{7ca}$\cdot$ \Bapp (right). Vertical arrows indicate main contributions to the electronic transition to the lowest bright state. Bottom panel: Aromatic {\bf d}, and aza-crown ether {\bf e}, regions of proton NMR spectra of compound \textbf{7ca} upon addition of barium perchlorate. The most important changes in chemical shift (in ppm) are highlighted. All the spectra were recorded at 500 MHz.}
\label{fig:nmr}
\end{figure}

\begin{figure}[htb!]
\centering
\includegraphics[width=0.99\textwidth]{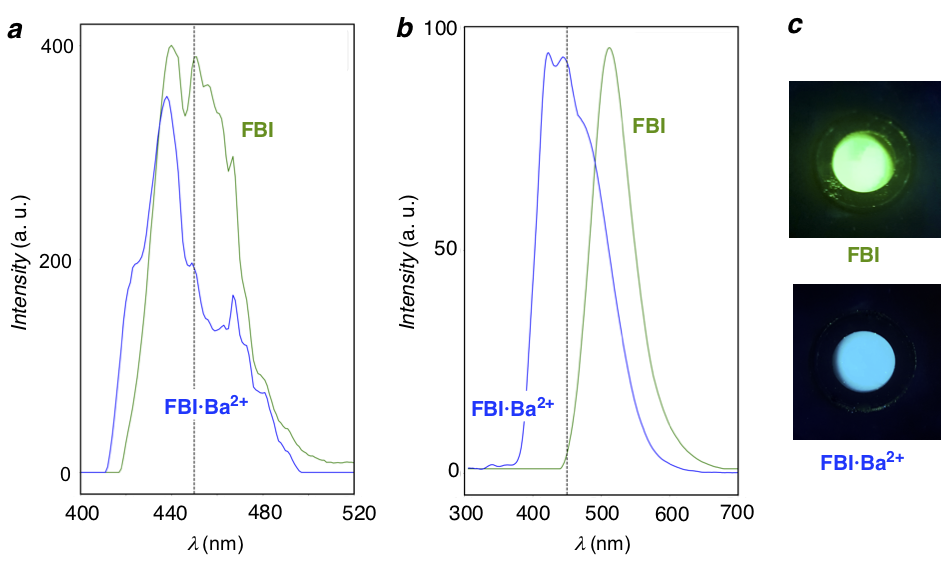}
\caption{{\it Response of FBI in two different dry media, PMMA and silica gel}. {\bf a}, Response of chelated and unchelated FBI indicators in PMMA, upon excitation at \SI{350}{nm}. 
{\bf b}, Emission spectra of FBI and FBI-\Bapp\ in silica powder, upon excitation at \SI{250}{nm}.  {\bf c}, Photographs of both species in silica, showing the characteristic bicolor emission observed in solution, upon irradiation at \SI{365}{nm}.}
\label{fig:poly}
\end{figure}

\begin{figure}[htb!]
\centering
\includegraphics[width=0.99\textwidth]{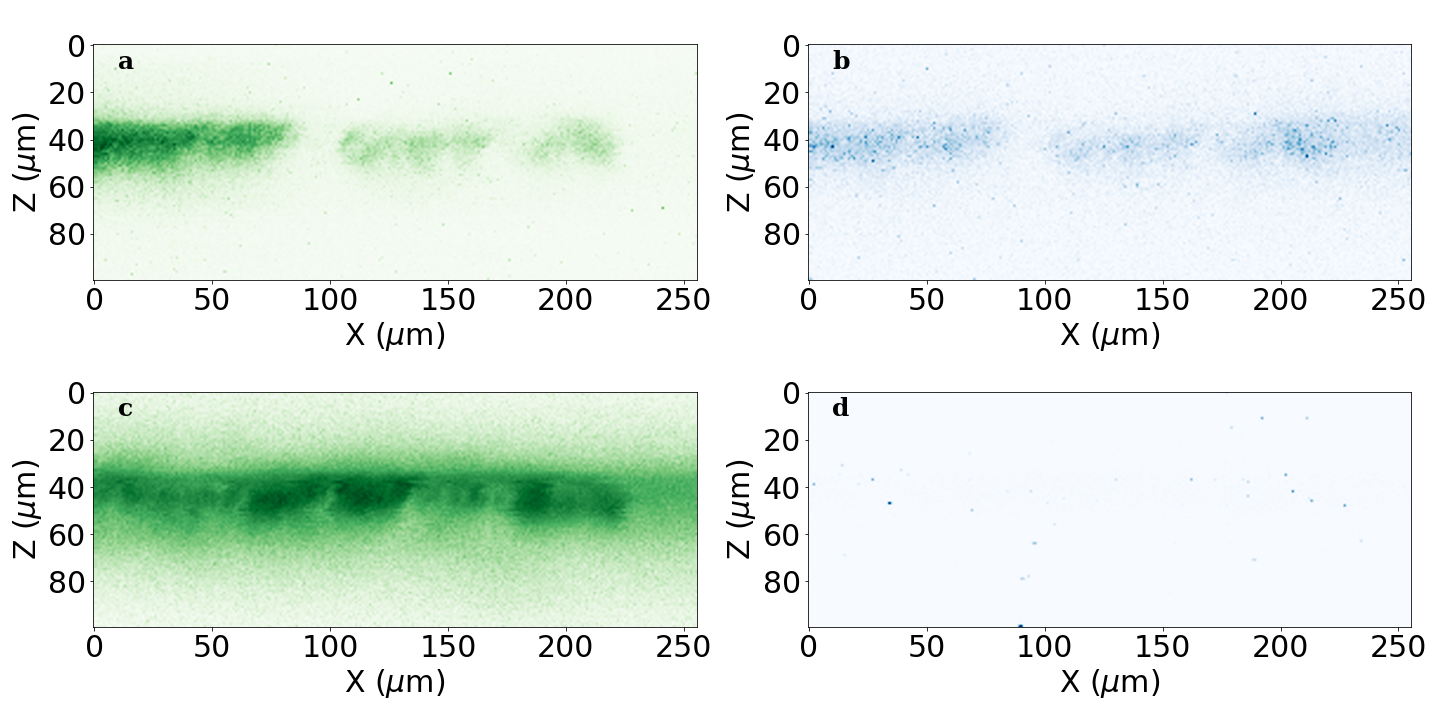} 
\caption{{\it Characterisation of FBI with TPA microscopy}: 
 TPA profiles on the SBFp and SFp pellets obtained with our \SI{800}{nm} laser. {\bf a}, SBFp applying a high-pass filter, $\lambda > 450$ nm (green panel). {\bf b}, SBFp applying a low-pass filter, $\lambda < 450$ nm (blue panel). {\bf c}, SFp applying a high-pass filter, $\lambda > 450$ nm (green panel). {\bf d}, SFp applying a low-pass filter, $\lambda < 450$ nm (blue panel). Notice that the green panel of the chelated indicators has sizeable fraction of the signal, while the blue panel of the unchelated indicators is essentially empty. }
\label{fig:laser}
\end{figure}


\clearpage

\section{Methods}

\subsection*{Laser setup}
A schematic diagram of our laser setup is depicted in Figure \ref{fig:espinardo}a.
We took advantage of the fact that
the emission spectra of FBI and FBI$\cdot$\Bapp for excitation light of \SI{250}{nm} and
excitation light of \SI{400}{nm} are very similar (Figure \ref{fig:espinardo}b), to use a mode-locked Ti:Sapphire infrared laser (\SI{800}{nm}) as illumination source, inducing the absorption of two photons of \SI{400}{nm} each. This laser system provided pulses of infrared light with a repetition rate of 76 MHz. The pulse duration was \SI{400}{fs} at the sample's plane.The beam was reflected at a dichroic mirror, passed a non-immersion objective (20x, A=0.5) and reached the sample, illuminating a spot limited by diffraction to a volume of about   \SI{1}{\cubic\micro\meter}. A DC-motor coupled to the objective allowed optical sectioning across the sample along the Z-direction. This image modality is known as XZ tomographic imaging \cite{Bueno2011}. We call {\it profiles} to those tomographic images (similar to the B scanning mode used in optical coherence tomography clinical devices). The emitted light was collected through the same objective and passed the dichroic mirror. Finally, before reaching the photomultiplier tube used as detection unit, the TPA signal passed through either a low-pass ($\lambda < 450$ nm) or a high-pass ($\lambda > 450$ nm) spectral filter. 

In order to estimate the absolute number of fluorescence photons emitted by the FBI indicator in a TPA scan, we first measured a reference sample of fluorescein suspended in PVA (FRS). 
Figure  \ref{fig:espinardo}c shows a log-log plot of the recorded PMT signal as a function of the laser power for FRS. As expected for TPA, the slope of the resulting straight line has a value near 2. Figure  \ref{fig:espinardo}d, shows
a profile taken on FRS at a power of \SI{80}{\milli\watt}. Identical profiles were taken on SBFp at a power of \SI{40}{\milli\watt}. This allowed the measurement of the ratio
$\delta_r = \delta_{sbfp} /\delta_{frs}$, which turned out to be $\delta_r = 17 \pm 4$, and therefore, 
$\delta_{FBI\cdot\Bapp} = 6.2 \pm 1.7 \times 10^2$ GM. The details of the measurement are discussed below.

\subsection*{Determination of the brightness of FBI relative to fluorescein}
The fluorophore brightness ($\delta = \sigma \cdot \phi_\lambda$, where $\sigma$ is the TPA cross section and 
$\phi_\lambda$ the quantum yield) 
of fluorescein is well known for a wavelength of \SI{800}{nm} \cite{Xu1995}: 
$\delta_{fluo} = 36 \pm 9.7 $ GM 
(1 GM$= 10^{-50}$ cm$^4 \cdot$second$/$ (photon$\cdot$molecule)). It is, therefore, possible to normalise the
brightness of FBI to that of fluorescein, using samples of known concentrations and measuring the response in our setup for identical profiles. To that end, we used a control sample of fluorescein suspended in PVA (FPVA), with  
a concentration of $n_{fpva} = 10^{13}$ molecules/cm$^3$ and compared it with our FBI-chelated pellet (SBFp), which had a concentration of $n_{sbfp} =2.2 \times 10^{17}$ molecules/cm$^3$. Profiles were
taken on FPVA at a power of \SI{500}{\milli\watt}. Identical profiles were taken on SBFp at a power of \SI{100}{\milli\watt}. 
The total signal integrated by the PMT in both the FPVA and SBFp samples is:

\begin{equation}
I = K \cdot n \cdot \delta \cdot P^2
\end{equation}
where $n$ is the density of molecules (molecules/cm$^3$) in the sample and $P$ is the laser power. $K$ is a constant which depends of the setup, but is the same for the FPVA and SBFp profiles. It follows that:

\begin{equation}
R_{fbi/fluo} = \frac{\delta_{sbfp}}{\delta_{fpva}} = \frac{I_{sbfp}}{I_{fpva}} \frac{n_{fpva}}{n_{sbfp}} (\frac{P_{fpva}}{P_{sbfp}})^2
\label{eq:r}
\end{equation}

All the quantities in equation \ref{eq:r} are known. In particular, the integral of the SBFp profile yields
$10^9$ PMT counts, while the integral of the FPVA profile results in $5.9 \times 10^4$ counts. Thus, we find :
$R_{fluo/fbi} = 17 \pm 4$, where the $\sim$ 20\% relative error is dominated by the uncertainty in the concentration 
$n_{sbfp}$, and therefore, $\delta_{FBI\cdot\Bapp} = 6.2 \pm 1.2 \times 10^2$ GM.

\subsection*{Interaction of FBI with other metals}

The interaction of FBI ({\bf 7ca}) with other metals in the same column of barium was studied in order to assess the
selectivity of the indicator. The results are summarised in Figure \ref{fig:ca}. Solutions ($5 \times 10^{-5}$ M) of {\bf 7ca} and metal source in ratio 1:1 were prepared for this study. We used Ca(OH)$_2$, Mg(ClO$_4$)$_2$, Sr(ClO$_4$)$_2$ and Ba(ClO$_4$)$_2$ with CH$_3$CN as a solvent. We observed that  compound {\bf 7ca} was capable to chelate Sr$^{2+}$ and Ba$^{2+}$. This was expected, given their similar atomic radii ($2.15$ \AA\ for Sr$^{2+}$ and $2.22$ \AA\  for Ba$^{2+}$). It follows that {\bf 7ca} should be able to chelate Ra$^{2+}$ (atomic radius of $2.2$ \AA). On the other hand, we observed that 
Ca$^{2+}$ (atomic radius of $1.97$ \AA) and Mg$^{2+}$ (atomic radius of $1.6$ \AA) were not chelated. It follows that the
lightest alkalyne earth, beryllium, will not be chelated.  



\clearpage
\appendix
\counterwithin{figure}{section}
\section{Extended data figures and tables}

\begin{figure}[h!]
\centering
\includegraphics[width=1.1\textwidth]{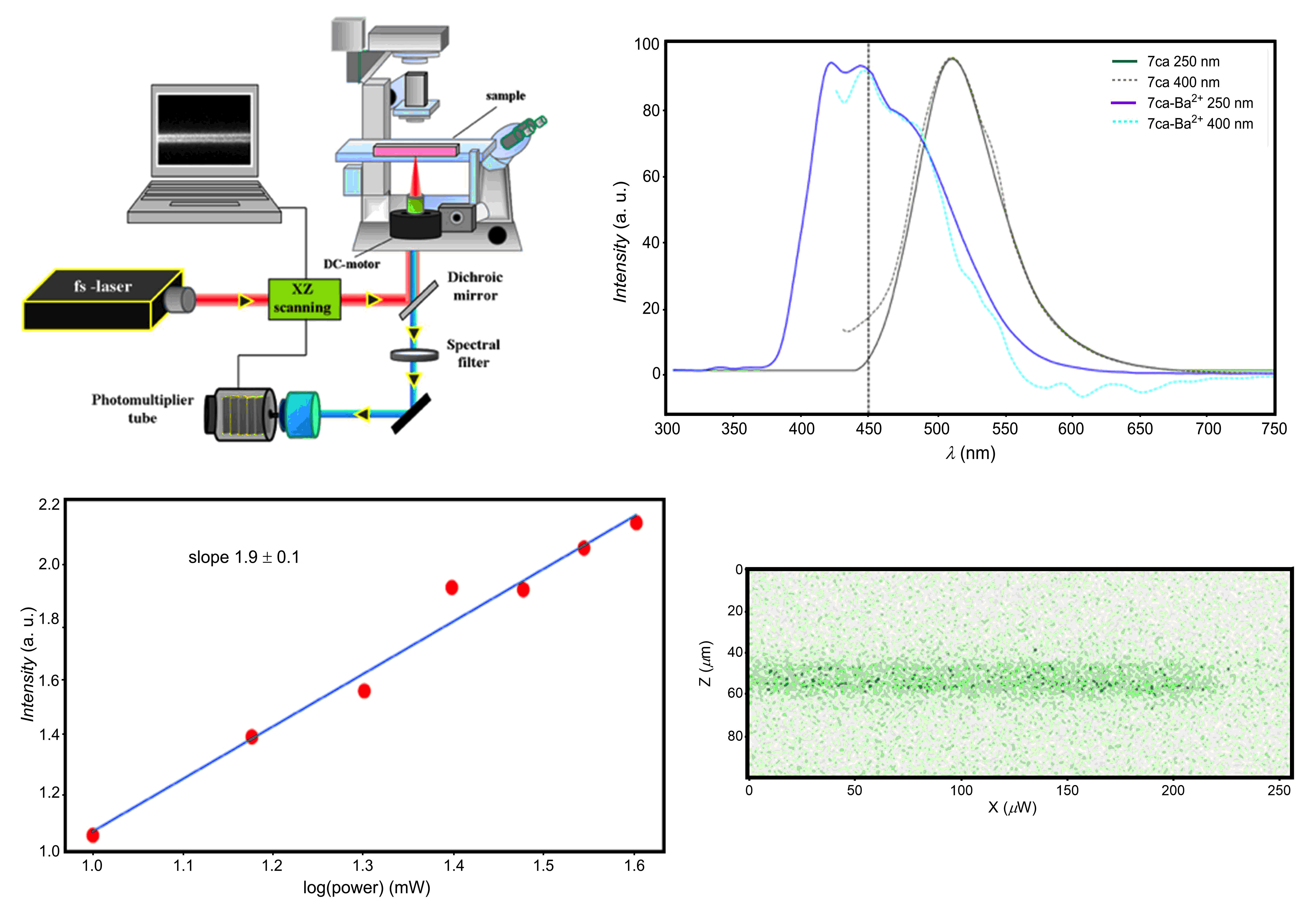} 
\caption{{\it TPA microscopy}. {\bf a}, A cartoon of our setup. An infrared (\SI{800}{nm}) laser passes through a dichroic and fills the back plane of the objective (20x NA = 0.5) of an inverted microscope. The laser is focused in the sample, with a spot limited by diffraction (\eg\ a volume of about \SI{1}{\cubic\micro\meter}). The emitted fluorescence passes through a selection filter before being recorded by a photomultiplier. {\bf b}, Emission spectra of FBI and FBI$\cdot$\Bapp for excitation light of \SI{250}{nm} (green, blue) and an
excitation light of \SI{400}{nm} (olive, cyan). The spectra are very similar, allowing the use of an
infrared laser of \SI{800}{nm} for our proof-of-concept.  {\bf c}, Log-log plot showing the quadratic dependence of the intensity with the power, characteristic of TPA, for a fluorescein reference sample (FRS). {\bf d}, Two-dimensional scan (profile) across the FRS. Integration of the profile yields and
integrated signal which can be used as normalisation for the FBI samples. }
\label{fig:espinardo}
\end{figure}

\begin{figure} [h!]
\centering
\includegraphics[width=0.99\textwidth]{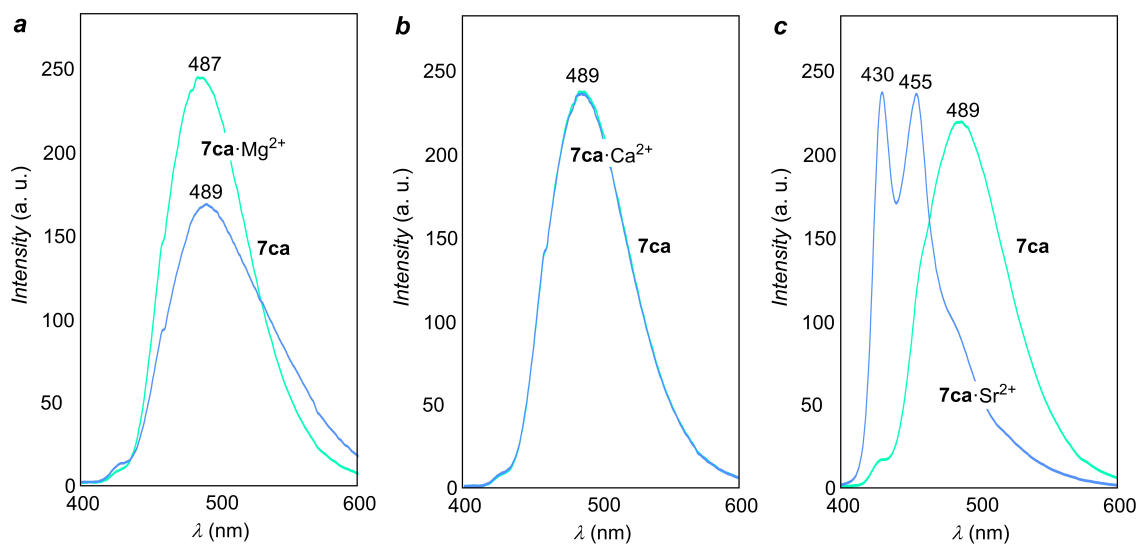} 
\caption{{\it Interaction of FIB with other metals}. {\bf a}, FIB$\cdot$Mg$^{2+}$ (blue) and unchelated (green) indicators.
{\bf b}, FIB$\cdot$Ca$^{2+}$ (blue) and unchelated indicators. {\bf c}, FIB$\cdot$Sr$^{2+}$ (blue) and unchelated (green) indicators. In the first two cases, the spectra show that FIB is not chelated with the ion, while in the third case
the response is similar to barium, showing the formation of a supramolecular complex. All excitation spectra taken at
\SI{250}{nm}. }
\label{fig:ca}
\end{figure}

\begin{figure} [h!]
\centering
\includegraphics[width=0.99\textwidth]{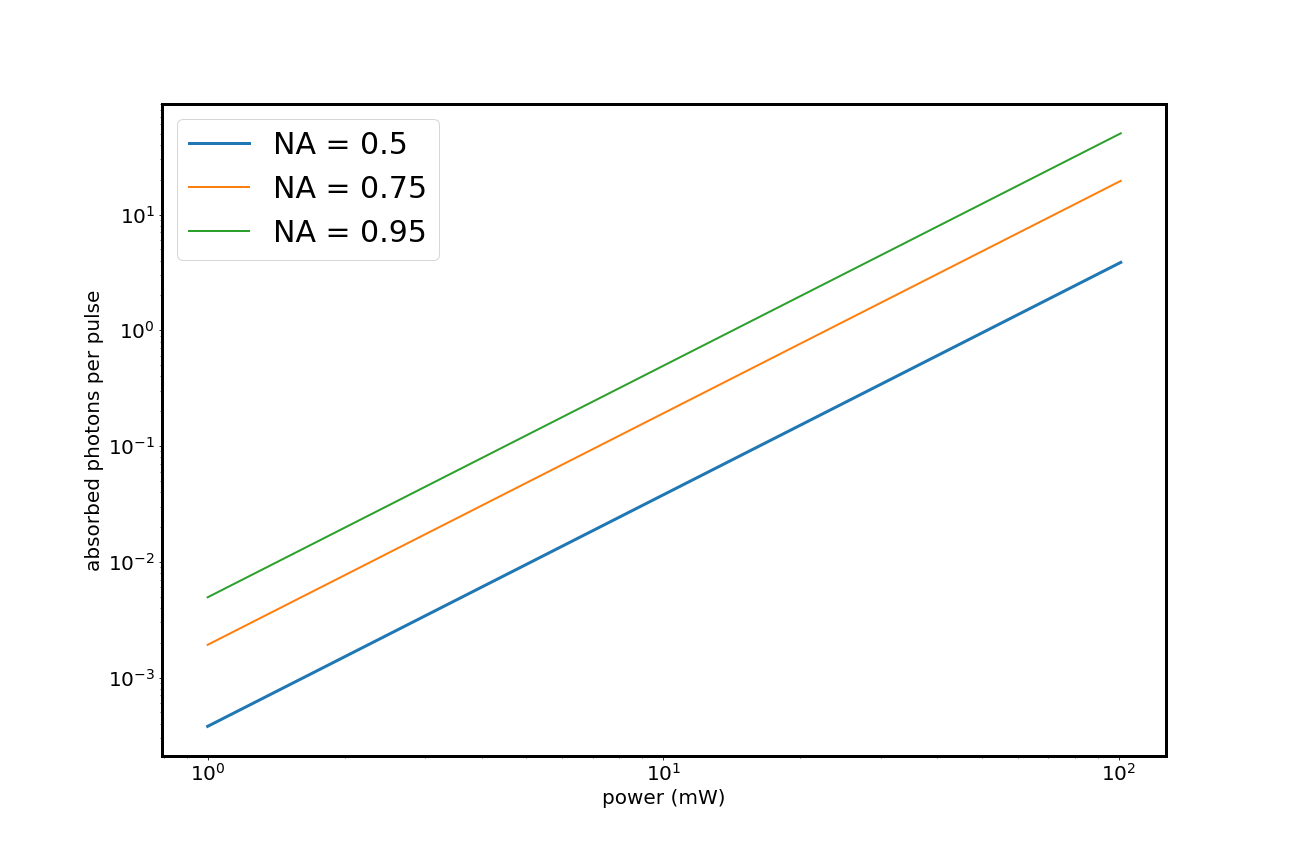} 
\caption{Number of absorbed photons per chelated fluorophere as a function of the laser power for different values
of the numerical aperture A. }
\label{fig:na}
\end{figure}

\clearpage

\acknowledgments
We acknowledge the support of our colleagues of the NEXT collaboration in the development of this work as a part of the R\&D program to to develop a background-free experiment based in \Bapp\ tagging. We also acknowledge 
support from the following agencies and institutions: the European Research Council (ERC) under the Advanced Grant 339787-NEXT; the Ministry of Science, Universities and Research of Spain and FEDER under grants FIS2014-53371-C04, 
FIS2016-76163-R, MINECO/FEDER CT2016-80955-P, CTQ2016-80375-P and CTQ2014-51912-REDC;  the Basque Government (GV/EJ, grant IT-324-07 and IT1180-19). Agencia de Ciencia y Tecnolog\'ia de la Regi\'on de Murcia (19897/GERM/15). The authors also thank the SGI/IZO-SGIker UPV/EHU, Fundaci\'on S\'eneca and the DIPC for generous allocation of computational and analytical resources. 

\textbf{Author contributions}. J.J.G.C., F.P.C. and D.N. conceived the project. J.J.G.C. and F.P.C. coordinated the experiments and analysed the data. I.R. and B.A. carried out the chemical synthesis, characterisation and solution fluorescence studies of the compounds. J.I.M. carried out the NMR experiments. C.T., F.P.C. and D.C. performed the computational studies. Z.F. carried out the silica experiments. B.O. and T.S. performed the solid phase experiments involving polymers. J.M.B., R.M.M., P.H., F.M. and P.A., performed the laser experiments (coordinated by J.M.B). J.J.G.C. and F.P.C. wrote the manuscript. D.N., F.M. and P.A. assisted in writing and editing the manuscript.

\textbf{Competing interests}. The authors declare no competing interests.

\bibliography{fibrefs}

\begin{thebibliography}{10}
\expandafter\ifx\csname url\endcsname\relax
  \def\url#1{\texttt{#1}}\fi
\expandafter\ifx\csname urlprefix\endcsname\relax\def\urlprefix{URL }\fi
\providecommand{\bibinfo}[2]{#2}
\providecommand{\eprint}[2][]{\url{#2}}

\bibitem{Majorana:1937}
\bibinfo{author}{Majorana, E.}
\newblock \bibinfo{title}{{Theory of the Symmetry of Electrons and Positrons}}.
\newblock \emph{\bibinfo{journal}{Nuovo Cim.}} \textbf{\bibinfo{volume}{14}},
  \bibinfo{pages}{171--184} (\bibinfo{year}{1937}).

\bibitem{Fukugita:1986hr}
\bibinfo{author}{Fukugita, M.} \& \bibinfo{author}{Yanagida, T.}
\newblock \bibinfo{title}{{Baryogenesis Without Grand Unification}}.
\newblock \emph{\bibinfo{journal}{Phys. Lett.}}
  \textbf{\bibinfo{volume}{B174}}, \bibinfo{pages}{45--47}
  (\bibinfo{year}{1986}).

\bibitem{Gando:2016ji}
\bibinfo{author}{Gando, A.} \emph{et~al.}
\newblock \bibinfo{title}{{Search for Majorana Neutrinos Near the Inverted Mass
  Hierarchy Region with KamLAND-Zen }}.
\newblock \emph{\bibinfo{journal}{Phys. Rev. Lett.}}
  \textbf{\bibinfo{volume}{117}}, \bibinfo{pages}{109903}
  (\bibinfo{year}{2016}).
\newblock
  \urlprefix\url{https://link.aps.org/doi/10.1103/PhysRevLett.117.109903}.

\bibitem{Agostini:2018tnm}
\bibinfo{author}{Agostini, M.} \emph{et~al.}
\newblock \bibinfo{title}{{Improved Limit on Neutrinoless Double-$\beta$ Decay
  of $^{76}$Ge from GERDA Phase II}}.
\newblock \emph{\bibinfo{journal}{Phys. Rev. Lett.}}
  \textbf{\bibinfo{volume}{120}}, \bibinfo{pages}{132503}
  (\bibinfo{year}{2018}).
\newblock \eprint{arXiv.1803.11100}.

\bibitem{Alduino:2017ehq}
\bibinfo{author}{Alduino, C.} \emph{et~al.}
\newblock \bibinfo{title}{{First Results from CUORE: A Search for Lepton Number
  Violation via $0\nu\beta\beta$ Decay of $^{130}$Te}}.
\newblock \emph{\bibinfo{journal}{Phys. Rev. Lett.}}
  \textbf{\bibinfo{volume}{120}}, \bibinfo{pages}{132501}
  (\bibinfo{year}{2018}).
\newblock \eprint{arXiv.1710.07988}.

\bibitem{Gomez-Cadenas:2019sfa}
\bibinfo{author}{Gomez-Cadenas, J.~J.}
\newblock \bibinfo{title}{{Status and prospects of the NEXT experiment for
  neutrinoless double beta decay searches}} (\bibinfo{year}{2019}).
\newblock \eprint{arXiv.1906.01743}.

\bibitem{Moe:1991ik}
\bibinfo{author}{Moe, M.~K.}
\newblock \bibinfo{title}{{New approach to the detection of neutrinoless double
  beta decay}}.
\newblock \emph{\bibinfo{journal}{Physical Review}}
  \textbf{\bibinfo{volume}{C44}}, \bibinfo{pages}{931--934}
  (\bibinfo{year}{1991}).

\bibitem{Danilov:2000pp}
\bibinfo{author}{Danilov, M.} \emph{et~al.}
\newblock \bibinfo{title}{{Detection of very small neutrino masses in double
  beta decay using laser tagging}}.
\newblock \emph{\bibinfo{journal}{Physics Letters}}
  \textbf{\bibinfo{volume}{B480}}, \bibinfo{pages}{12--18}
  (\bibinfo{year}{2000}).
\newblock \eprint{hep-ex/0002003}.

\bibitem{Sinclair:2011zz}
\bibinfo{author}{Sinclair, D.} \emph{et~al.}
\newblock \bibinfo{title}{{Prospects for Barium Tagging in Gaseous Xenon}}.
\newblock \emph{\bibinfo{journal}{Journal of Physics Conference Series}}
  \textbf{\bibinfo{volume}{309}}, \bibinfo{pages}{012005}
  (\bibinfo{year}{2011}).

\bibitem{Mong:2014iya}
\bibinfo{author}{Mong, B.} \emph{et~al.}
\newblock \bibinfo{title}{{Spectroscopy of Ba and Ba$^+$ deposits in solid
  xenon for barium tagging in nEXO}}.
\newblock \emph{\bibinfo{journal}{Physical Review}}
  \textbf{\bibinfo{volume}{A91}}, \bibinfo{pages}{022505}
  (\bibinfo{year}{2015}).
\newblock \eprint{arXiv.1410.2624}.

\bibitem{Chambers:2018srx}
\bibinfo{author}{Chambers, C.} \emph{et~al.}
\newblock \bibinfo{title}{{Imaging individual barium atoms in solid xenon for
  barium tagging in nEXO}}.
\newblock \emph{\bibinfo{journal}{Nature}} \textbf{\bibinfo{volume}{569}},
  \bibinfo{pages}{203--207} (\bibinfo{year}{2019}).
\newblock \eprint{arXiv.1806.10694}.

\bibitem{PhysRevC.92.045504}
\bibinfo{author}{Albert, J.~B.} \emph{et~al.}
\newblock \bibinfo{title}{Measurements of the ion fraction and mobility of
  alpha- and beta-decay products in liquid xenon using the exo-200 detector}.
\newblock \emph{\bibinfo{journal}{Phys. Rev. C}} \textbf{\bibinfo{volume}{92}},
  \bibinfo{pages}{045504} (\bibinfo{year}{2015}).
\newblock \urlprefix\url{http://link.aps.org/doi/10.1103/PhysRevC.92.045504}.

\bibitem{1997NIMPA.396..360B}
\bibinfo{author}{{Bolotnikov}, A.} \& \bibinfo{author}{{Ramsey}, B.}
\newblock \bibinfo{title}{{The spectroscopic properties of high-pressure
  xenon}}.
\newblock \emph{\bibinfo{journal}{Nuclear Instruments and Methods in Physics
  Research A}} \textbf{\bibinfo{volume}{396}}, \bibinfo{pages}{360--370}
  (\bibinfo{year}{1997}).

\bibitem{Nygren_2015}
\bibinfo{author}{Nygren, D.~R.}
\newblock \bibinfo{title}{Detecting the barium daughter in136xe
  0-$\nu\beta\beta$ decay using single-molecule fluorescence imaging
  techniques}.
\newblock \emph{\bibinfo{journal}{Journal of Physics: Conference Series}}
  \textbf{\bibinfo{volume}{650}}, \bibinfo{pages}{012002}
  (\bibinfo{year}{2015}).

\bibitem{Nygren:2009zz}
\bibinfo{author}{Nygren, D.}
\newblock \bibinfo{title}{{High-pressure xenon gas electroluminescent TPC for
  $0-\nu ~ \beta \beta$-decay search}}.
\newblock \emph{\bibinfo{journal}{Nucl.Instrum.Meth.}}
  \textbf{\bibinfo{volume}{A603}}, \bibinfo{pages}{337--348}
  (\bibinfo{year}{2009}).

\bibitem{Alvarez:2012haa}
\bibinfo{author}{\'Alvarez, V.} \emph{et~al.}
\newblock \bibinfo{title}{{NEXT-100 Technical Design Report (TDR): Executive
  Summary}}.
\newblock \emph{\bibinfo{journal}{JINST}} \textbf{\bibinfo{volume}{7}},
  \bibinfo{pages}{T06001} (\bibinfo{year}{2012}).
\newblock \eprint{arXiv.1202.0721}.

\bibitem{Martin-Albo:2015rhw}
\bibinfo{author}{Mart\'in-Albo, J.} \emph{et~al.}
\newblock \bibinfo{title}{{Sensitivity of NEXT-100 to Neutrinoless Double Beta
  Decay}}.
\newblock \emph{\bibinfo{journal}{JHEP}} \textbf{\bibinfo{volume}{05}},
  \bibinfo{pages}{159} (\bibinfo{year}{2016}).
\newblock \eprint{arXiv.1511.09246}.

\bibitem{Jones:2016qiq}
\bibinfo{author}{Jones, B. J.~P.}, \bibinfo{author}{McDonald, A.~D.} \&
  \bibinfo{author}{Nygren, D.~R.}
\newblock \bibinfo{title}{{Single Molecule Fluorescence Imaging as a Technique
  for Barium Tagging in Neutrinoless Double Beta Decay}}.
\newblock \emph{\bibinfo{journal}{JINST}} \textbf{\bibinfo{volume}{11}},
  \bibinfo{pages}{P12011} (\bibinfo{year}{2016}).
\newblock \eprint{arXiv.1609.04019}.

\bibitem{Renner:2019pfe}
\bibinfo{author}{Renner, J.} \emph{et~al.}
\newblock \bibinfo{title}{{Energy calibration of the NEXT-White detector with
  1\% resolution near Q$\beta\beta$ of $^{136}$Xe}}  (\bibinfo{year}{2019}).
\newblock \eprint{arXiv.1905.13110}.

\bibitem{Ferrario:2019kwg}
\bibinfo{author}{Ferrario, P.} \emph{et~al.}
\newblock \bibinfo{title}{{Efficiency of the topological signature in the
  NEXT-White detector}}  (\bibinfo{year}{2019}).
\newblock \eprint{arXiv.1905.13141}.

\bibitem{Bainglass:2018odn}
\bibinfo{author}{Bainglass, E.}, \bibinfo{author}{Jones, B. J.~P.},
  \bibinfo{author}{Foss, F.~W.}, \bibinfo{author}{Huda, M.~N.} \&
  \bibinfo{author}{Nygren, D.~R.}
\newblock \bibinfo{title}{{Mobility and Clustering of Barium Ions and Dications
  in High Pressure Xenon Gas}}.
\newblock \emph{\bibinfo{journal}{Phys. Rev.}} \textbf{\bibinfo{volume}{A97}},
  \bibinfo{pages}{062509} (\bibinfo{year}{2018}).
\newblock \eprint{arXiv.1804.01169}.

\bibitem{ARAI201456}
\bibinfo{author}{Arai, F.} \emph{et~al.}
\newblock \bibinfo{title}{Investigation of the ion surfing transport method
  with a circular rf carpet}.
\newblock \emph{\bibinfo{journal}{International Journal of Mass Spectrometry}}
  \textbf{\bibinfo{volume}{362}}, \bibinfo{pages}{56 -- 58}
  (\bibinfo{year}{2014}).
\newblock
  \urlprefix\url{http://www.sciencedirect.com/science/article/pii/S1387380614000098}.

\bibitem{Jeong:2012}
\bibinfo{author}{Jeong, Y.} \& \bibinfo{author}{Yoon, J.}
\newblock \bibinfo{title}{Recent progress on fluorescent chemosensors for metal
  ions.}
\newblock \emph{\bibinfo{journal}{Inorg. Chim. Acta, 381, 2-14}}
  (\bibinfo{year}{2012}).

\bibitem{Carter:2014}
\bibinfo{author}{Carter, K.~P.}, \bibinfo{author}{Young, A.~M.} \&
  \bibinfo{author}{Palmer, A.~E.}
\newblock \bibinfo{title}{Fluorescent sensors for measuring metal ions in
  living systems.}
\newblock \emph{\bibinfo{journal}{Chem. Rev. (Washington, DC, U. S.) 114 (8),
  4564-4601.}}  (\bibinfo{year}{2014}).

\bibitem{Thapa:2019zjk}
\bibinfo{author}{Thapa, P.} \emph{et~al.}
\newblock \bibinfo{title}{{Barium Chemosensors with Dry-Phase Fluorescence for
  Neutrinoless Double Beta Decay}}  (\bibinfo{year}{2019}).
\newblock \eprint{arXiv.1904.05901}.

\bibitem{McDonald:2017izm}
\bibinfo{author}{McDonald, A.~D.} \emph{et~al.}
\newblock \bibinfo{title}{{Demonstration of Single Barium Ion Sensitivity for
  Neutrinoless Double Beta Decay using Single Molecule Fluorescence Imaging}}.
\newblock \emph{\bibinfo{journal}{Phys. Rev. Lett.}}
  \textbf{\bibinfo{volume}{120}}, \bibinfo{pages}{132504}
  (\bibinfo{year}{2018}).
\newblock \eprint{arXiv.1711.04782}.

\bibitem{Aginagalde:2010}
\bibinfo{author}{Aginagalde, M.} \emph{et~al.}
\newblock \bibinfo{title}{Tandem [8 + 2] cycloaddition-[2 + 6 + 2]
  dehydrogenation reactions involving imidazo[1,2-a]pyridines and
  imidazo[1,2-a]pyrimidines.}
\newblock \emph{\bibinfo{journal}{J. Org. Chem., 75 (9), 2776-2784}}
  (\bibinfo{year}{2010}).

\bibitem{Zhang:2018}
\bibinfo{author}{Zhang, Y.}, \bibinfo{author}{Tang, S.},
  \bibinfo{author}{Thapaliya, E.~R.}, \bibinfo{author}{Sansalone, L.} \&
  \bibinfo{author}{Raymo, F.~M.}
\newblock \bibinfo{title}{Fluorescence activation with switchable oxazines.}
\newblock \emph{\bibinfo{journal}{Chem. Commun. (Cambridge, U. K.), 54 (64),
  8799-8809.}}  (\bibinfo{year}{2018}).

\bibitem{Ko:2018}
\bibinfo{author}{Ko, C.-C.} \& \bibinfo{author}{Yam, V. W.-W.}
\newblock \bibinfo{title}{Coordination compounds with photochromic ligands:
  Ready tunability and visible light-sensitized photochromism.}
\newblock \emph{\bibinfo{journal}{Acc. Chem. Res. , 51 (1), 149-159}}
  (\bibinfo{year}{2018}).

\bibitem{Maitra:2017}
\bibinfo{author}{Maitra, R.}, \bibinfo{author}{Chen, J.-H.},
  \bibinfo{author}{Hu, C.-H.} \& \bibinfo{author}{Lee, H.~M.}
\newblock \bibinfo{title}{Synthesis and optical properties of push-push-pull
  chromophores based on imidazo[5,1,2-cd]indolizines and
  naphtho[1',2':4,5]imidazo[1,2-a]pyridines.}
\newblock \emph{\bibinfo{journal}{Eur. J. Org. Chem. (40), 5975-5985}}
  (\bibinfo{year}{2017}).

\bibitem{Benesi:1949}
\bibinfo{author}{Benesi, H.~A.} \& \bibinfo{author}{Hildebrand, J.~H.}
\newblock \bibinfo{title}{A spectrophotometric investigation of the interaction
  of iodine with aromatic hydrocarbons.}
\newblock \emph{\bibinfo{journal}{J. Am. Chem. Soc., 71, 2703--2707}}
  (\bibinfo{year}{1949}).

\bibitem{ZhangDuan:2018}
\bibinfo{author}{Zhang, Q.} \& \bibinfo{author}{Duan, K.}
\newblock \bibinfo{title}{Fluorescence chemosensor containing
  4-methyl-7-coumarinyloxy, acetylhydrazono and n-phenylaza-15-crown-5 moieties
  for k+ and ba2+ ions.}
\newblock \emph{\bibinfo{journal}{Heterocycl. Commun., 24 (3), 141--145}}
  (\bibinfo{year}{2018}).

\bibitem{Batsanov:2001}
\bibinfo{author}{Batsanov, S.~S.}
\newblock \bibinfo{title}{Van der waals radii of elements.}
\newblock \emph{\bibinfo{journal}{Inorganic Materials, 37 (9), 871--895}}
  (\bibinfo{year}{2001}).

\bibitem{Denk990}
\bibinfo{author}{W~Denk, W.~W., JH~Strickler}.
\newblock \bibinfo{title}{Two-photon laser scanning fluorescence microscopy}.
\newblock \emph{\bibinfo{journal}{Science, Vol. 248, Issue 4951, pp. 73-76 DOI:
  10.1126/science.2321027}}  (\bibinfo{year}{990}).

\bibitem{Avila2019}
\bibinfo{author}{Ávila, F.~J.}, \bibinfo{author}{Gambín, A.},
  \bibinfo{author}{P., A.} \& \bibinfo{author}{M., B.~J.}
\newblock \bibinfo{title}{In vivo two-photon microscopy of the human eye}.
\newblock \emph{\bibinfo{journal}{Sci. Reports 9, 10121}}
  (\bibinfo{year}{2019}).

\bibitem{Bueno2011}
\bibinfo{author}{Bueno, J.~M.} \emph{et~al.}
\newblock \bibinfo{title}{Multiphoton microscopy of ex-vivo corneas after
  collagen cross-linking}.
\newblock \emph{\bibinfo{journal}{Invest. Ophthalmol. Vis. Sci. 52(8),
  5325-5331}}  (\bibinfo{year}{2011}).

\bibitem{Xu1995}
\bibinfo{author}{Xu, C.} \& \bibinfo{author}{Webb, W.~W.}
\newblock \bibinfo{title}{Measurement of two-photon excitation cross sections
  of molecular fluorophores with data from 690 to 1050 nm}.
\newblock \emph{\bibinfo{journal}{J. Opt. Soc. Am. B 481}}
  (\bibinfo{year}{1995}).

\end{thebibliography}

\end{document}